# Efficient Spin-Orbit Torque Switching with Non-Epitaxial Chalcogenide Heterostructures

Tian-Yue Chen[1], Cheng-Wei Peng[1], Tsung-Yu Tsai[1], Wei-Bang Liao[1], Chun-Te Wu[1], Hung-Wei Yen[1,3], and Chi-Feng Pai[1,2*]

[1]Department of Materials Science and Engineering, National Taiwan University, Taipei 10617, Taiwan

[2]Center of Atomic Initiative for New Materials, National Taiwan University, Taipei 10617, Taiwan

[3]Advanced Research Center of Green Materials Science & Technology, National Taiwan University, Taipei 10617, Taiwan

**ABSTRACT:** The spin-orbit torques (SOTs) generated from topological insulators (TIs) have gained increasing attention in recent years. These TIs, which are typically formed by epitaxially grown chalcogenides, possess extremely high SOT efficiencies and have great potential to be employed in the next-generation spintronics devices. However, epitaxy of these chalcogenides is required to ensure the existence of topologically-protected surface state (TSS), which limits the feasibility of using these materials in industry. In this work, we show that non-epitaxial $Bi_xTe_{1-x}$/ferromagnet heterostructures prepared by conventional magnetron sputtering possess giant SOT efficiencies even without TSS. Through harmonic voltage measurement and hysteresis loop shift measurement, we find that the damping-like SOT efficiencies originated from the bulk spin-orbit interactions of such non-epitaxial heterostructures can reach values greater than 100% at room temperature. We further demonstrate current-induced SOT switching in these $Bi_xTe_{1-x}$-based heterostructures with thermally stable ferromagnetic layers, which indicates that such non-epitaxial chalcogenide materials can be potential efficient SOT sources in future SOT magnetic memory devices.

[*] Email: cfpai@ntu.edu.tw

## 1. INTRODUCTION

Topological insulators (TIs) have become a new family of materials systems for various spintronics applications [1-6]. TIs are bulk insulated while electrons are conducted through surface state [1, 7-8]. The surface state provides strong spin momentum locking and thus can be effectively used to achieve high spin-orbit torque (SOT) efficiency. Current SOT devices mostly utilize heavy metals (HMs) like Ta [9], W [10], Pt [11]…etc as spin current sources and the SOT efficiencies from the spin-Hall effect (SHE) in these conventional materials typically ranges from 0.1 to 0.3, which are still not large enough to satisfy the requirements for memory devices. Therefore, TIs are potential candidates to replace HMs in the next generation magnetic random access memory (MRAM) devices [12] due to their large SOT efficiencies. However, although TIs have been reported to possess greater charge-to-spin conversion efficiencies while compared to HMs, the preparation process for such materials systems in well-texture crystalline phase is still a challenging task. Among all materials, crystalline chalcogenides grown by molecular beam epitaxy (MBE) or ultra-high vacuum sputtering such as $Bi_xSe_{1-x}$ [13-16], $Bi_xSb_{1-x}$ [4], $Bi_xTe_{1-x}$ [2, 17] are discovered to possess topological insulating properties. Although recent TI-spintronics research have been focusing on $Bi_xSe_{1-x}$ and $Bi_xSb_{1-x}$ families as they both possess high SOT efficiencies, $Bi_xTe_{1-x}$ is also a potential candidate for spintronics device applications. Due to that, we choose sputter-deposited $Bi_xTe_{1-x}$ as our SOT source material to study its charge-to-spin conversion property.

In this work, we characterize the damping-like (DL) SOT efficiencies from various $Bi_xTe_{1-x}$-based magnetic heterostructures. We use Pt/Co/Pt trilayer structure as the effective ferromagnetic layer in adjacent to $Bi_xTe_{1-x}$. This symmetric trilayer structure yields a magnetization easy axis perpendicular to the film plane and possesses suitable out-of-plane coercivity to carry out the subsequent electrical measurements [18-20]. By performing harmonic voltage measurement [21] and hysteresis loop shift measurement [22] on $Bi_xTe_{1-x}$/Pt/Co/Pt heterostructures, the estimated DL-SOT efficiency of the as-deposited non-epitaxial $Bi_xTe_{1-x}$ can reach as high as $|\xi_{DL}| \approx 2.43$ (harmonic method) or $|\xi_{DL}| \approx 1.1$ (loop-shfit method) at room temperature. The post-annealing $Bi_xTe_{1-x}$ with polycrystalline structure also generates giant DL-SOT efficiency, $|\xi_{DL}| \approx 2.25$ (loop-shift method). Current-induced SOT switching measurements using these non-epitaxial $Bi_xTe_{1-x}$-based heterostructures are also performed, from which the giant DL-SOT efficiency $|\xi_{DL}| \geq 100\%$ is again verified. The origin of the observed giant charge-to-spin conversion efficiency in these non-epitaxial chalcogenides is further tentatively attributed to the bulk spin-orbit interaction therein.

## 2. EXPERIMENTAL RESULTS AND DISCUSSION

The $Bi_xTe_{1-x}$($t_{Bi_xTe_{1-x}}$)/Pt(2)/Co(0.5)/Pt(2) (units in nanometers) heterostructures in this work are grown by magnetron sputtering with base pressure around $3 \times 10^{-8}$ Torr. Two kinds of samples are prepared, which are as-deposited and post-annealing $Bi_xTe_{1-x}$ thin films, respectively. The $Bi_xTe_{1-x}$ layer is sputtered from a $Bi_2Te_3$ target, whereas the Pt and Co layers are sputtered from pure Pt and Co targets, respectively. Note that for the post-annealing samples, the $Bi_xTe_{1-x}$ layer is first deposited then annealed at 300 °C without breaking vacuum. The Pt/Co/Pt trilayers are then in-situ deposited onto the annealed $Bi_xTe_{1-x}$ layers after cooling down to room temperature. The Pt/Co/Pt symmetric structure has a robust perpendicular magnetic anisotropy (PMA), which is necessary for both harmonic voltage measurement and hysteresis loop shift measurement. The chalcogenide layer thickness $t_{Bi_xTe_{1-x}}$ ranges from 5 nm to 25 nm, and the thickness is verified by atomic force microscopy (AFM). To confirm the composition of our sputtered $Bi_xTe_{1-x}$ thin film, we use electron probe X-ray microanalyzer (EPMA) to confirm the atomic ratio between Bi and Te, which is close to 1:1 (See supplementary material S1). Therefore, in this work we use $Bi_xTe_{1-x}$ to represent our sputter-deposited $Bi_xTe_{1-x}$ film, where $x \approx 0.5$. We then examine the structural properties of the sputtered $Bi_xTe_{1-x}$ films by high-resolution transmission electron microscope (HR-TEM) as shown in Figure 1a, b. From the cross-section TEM image of a 15 nm as-deposited and a 15 nm post-annealing $Bi_xTe_{1-x}$ thin film, the as-deposited film is predominately amorphous and partially crystalline near the substrate, while the post-annealing film shows poly-crystalline structure. Note that due to the overall non-epitaxial structure of our deposited films, there should be no topologically-protected surface states (TSS) in our $Bi_xTe_{1-x}$ layer. Representative out-of-plane hysteresis loops obtained from an as-deposited and a post-annealing $Bi_xTe_{1-x}(15)/Pt(2)/Co(0.5)/Pt(2)$ heterostructure are shown in Figure 1c, d. The sharp hysteresis loops from both samples indicate that the easy axis of Co is indeed perpendicular to the film plane for both cases. The saturation magnetization of the Co layer is determined by vibrating sample magnetometry (VSM) to be $M_s = 1414 \pm 120$ emu/cm$^3$. The VSM result also suggests that there is no magnetic dead layer in the Pt/Co/Pt trilayer structure.

In order to characterize the SOT-induced effective fields acting upon the Co layer generated from $Bi_xTe_{1-x}$, we first perform harmonic voltage measurements on micron-sized Hall-bar devices patterned from an as-deposited $Bi_xTe_{1-x}(15)/Pt(2)/Co(0.5)/Pt(2)$ film. As illustrated in Figure 2a, an ac current with amplitude $I_{ac} = 2$ mA and frequency $f = 1000$ Hz is applied along $x$-direction while sweeping an in-plane magnetic field along either longitudinal ($H_x$) or transverse ($H_y$) direction. The first ($V_{1\omega}$) and the second ($V_{2\omega}$) harmonic Hall voltages are simultaneously recorded as functions of the sweeping fields. Representative $V_{1\omega}$ vs. $H_x$ results for both Co magnetization-up and magnetization-down states from an as-deposited $Bi_xTe_{1-x}(15)/Pt(2)/Co(0.5)/Pt(2)$ Hall-bar device are shown in Figure 2b. Figure 2c depicts representative $V_{2\omega}$ vs. $H_x$ and $V_{2\omega}$ vs. $H_y$ results for magnetization-up state obtained from the same device. The observed second harmonic Hall voltages are linearly proportional to the applied magnetic fields. The DL and field-like (FL) SOT-induced effective fields then can be evaluated from the first and second harmonic signals through [21, 23]

$$\Delta H_{DL(FL)} = -2 \frac{\partial V_{2\omega}/\partial H_{x(y)}}{\partial^2 V_{\omega}/\partial H_{x(y)}^2}. \tag{1}$$

Furthermore, the apparent DL(FL)-SOT efficiency can be calculated by [24]

$$\tilde{\xi}_{DL(FL)} = \frac{2e}{\hbar} \mu_0 M_s t_{Co} w t_{Bi_xTe_{1-x}} \left( \frac{\rho_{Pt/Co/Pt} t_{Bi_xTe_{1-x}} + \rho_{Bi_xTe_{1-x}} t_{Pt/Co/Pt}}{\rho_{Pt/Co/Pt} t_{Bi_xTe_{1-x}}} \right) \left( \frac{\Delta H_{DL(FL)}}{I_{ac}} \right), \tag{2}$$

where $\rho_{Bi_xTe_{1-x}} \approx 520\ \mu\Omega\text{-cm}$ and $\rho_{Pt/Co/Pt} \approx 25\ \mu\Omega\text{-cm}$ are the resistivities of the $Bi_xTe_{1-x}$ layer and the Pt/Co/Pt trilayer, respectively (See supplementary material S2 for layer resistivity characterization). $w = 5\ \mu m$ is the Hall-bar width. Since the $Bi_xTe_{1-x}$ layer is not directly in contact with the Co layer and the spins have to diffuse from $Bi_xTe_{1-x}$ to Co through the Pt(2) layer, we have to further take this spin transmission factor into account. The intrinsic SOT efficiencies of the $Bi_xTe_{1-x}$ layer is therefore estimated from the apparent values by $\xi_{DL(FL)} = \tilde{\xi}_{DL(FL)} \times \left( \text{sech}(t_{Pt}/\lambda_s^{Pt}) \right)^{-1}$, where $t_{Pt} = 2$ nm is the Pt layer thickness and $\lambda_s^{Pt} \approx 1.1$ nm is the Pt spin diffusion length (See supplemeantary material S3 for details). The intrinsic DL-SOT and FL-SOT efficiencies stemming from as-deposited $Bi_xTe_{1-x}(15)$ are found to be $\xi_{DL} \approx -2.43$ and $\xi_{FL} \approx -0.42$, respectively. We first note that the FL-SOT efficiency, though not negligible, is much smaller than the DL-SOT efficiency. Second, the estimated room temperature DL-SOT efficiency of $|\xi_{DL}| \sim 2.4$ from our as-deposited non-epitaxial $Bi_xTe_{1-x}$ is comparable to a recently reported value from a MBE-grown epitaxial chalcogenide (BiSb)$_2$Te$_3$ ($|\xi_{DL}|$ ~ 2.66) [25], which is characterized via a similar harmonic voltage approach. However, a recent study also points out that thermal contribution such as the ordinary Nernst effect can possibly lead to an overestimation of $|\xi_{DL}|$ if harmonic measurement is employed to characterize SOTs from chalcogenide systems [26]. Therefore, in order to characterize DL-SOT more directly, we further perform hysteresis loop shift and current-induced switching measurements.

We first verify the large DL-SOT efficiencies from both as-deposited and post-annealing Bi$_x$Te$_{1-x}$-based Hall-bar devices through hysteresis loop shift measurement. This DL-SOT characterization approach allows for a more direct probing of the SOT-caused current-induced effective fields without further interpretation on the detected electrical signals[11]. As shown in Figure 3a, a dc current $I_{dc}$ is injected into the above-mentioned micron-sized Hall-bar device with the application of an external in-plane bias field $H_x$. The in-plane field $H_x$ is required to break the interfacial Dzyaloshinskii-Moriya interaction (DMI) induced effective field and to realign the domain wall moments to facilitate magnetic domain expansion and domain wall propagation in the Co layer [27-28]. As shown in Figure 3b, a current-induced effective field $H_z^{eff}$ originated from the DL-SOT of as-deposited Bi$_x$Te$_{1-x}$(15) is produced when $I_{dc}$ is applied, which further causes a shift in the out-of-plane hysteresis loop. According to the slope of $H_z^{eff}$ vs. $I_{dc}$ plot as shown in the inset of Figure 3b, we can again estimate the apparent DL-SOT efficiency $\tilde{\xi}_{DL}$ of our samples. By further considering the current distribution due to the difference in resistivities between Bi$_x$Te$_{1-x}$ and Pt/Co/Pt layers, $\tilde{\xi}_{DL}$ is estimated through [29-30]

$$\tilde{\xi}_{DL} = \frac{2e}{\hbar}\left(\frac{2}{\pi}\right)\mu_0 M_s t_{Co} w t_{Bi_xTe_{1-x}} \left( \frac{\rho_{Pt/Co/Pt} t_{Bi_xTe_{1-x}} + \rho_{Bi_xTe_{1-x}} t_{Pt/Co/Pt}}{\rho_{Pt/Co/Pt} t_{Bi_xTe_{1-x}}} \right) \left( \frac{H_z^{eff}}{I_{dc}} \right). \quad (3)$$

In addition, we prepare control samples and confirm that the symmetric Pt(2)/Co(0.5)/Pt(2) trilayer has little SOT contribution on the whole heterostructure. In Figure 3c, the current-induced effective fields per current $H_z^{eff}/I_{dc}$ (as functions of $H_x$) for various control and experimental samples are presented. As expected, the magnitude of $H_z^{eff}/I_{dc}$ from the symmetric Pt(2)/Co(0.5)/Pt(2) structure is much smaller than those from $Bi_xTe_{1-x}$(15)/Pt(2)/Co(0.5)/Pt(2), W(3)/Pt(2)/Co(0.5)/Pt(2) (denoted as W(3)), and Pt(5)/Co(0.5)/Pt(2) (denoted as Pt(3)) control samples (See more details regarding control samples in Supporting Information S4). Note that the sign of $H_z^{eff}/I_{dc}$ under positive $H_x$ for both as-deposited and post-annealing $Bi_xTe_{1-x}$ are the same as that for W (opposite to that of Pt), which suggests a negative spin Hall ratio for $Bi_xTe_{1-x}$. Also note that most TI-related works report that epitaxial Bi-based chalcogenides have positive spin Hall ratios, which is opposite to our result[3-4, 15, 31]. This sign discrepancy is not surprising at all if one recall that the bulk spin-orbit interaction or the SHE have a rich variation in sign, depending on the carrier type of the materials system[32].

Again, the spin transmission factor has to be considered and the intrinsic DL-SOT efficiency of $Bi_xTe_{1-x}$ is further estimated by $\xi_{DL} = \tilde{\xi}_{DL} \times \left(\text{sech}\left(t_{Pt}/\lambda_s^{Pt}\right)\right)^{-1}$. The estimated $Bi_xTe_{1-x}$ thicknesses dependence of $|\xi_{DL}|$ from as-deposited and post-annealing devices are shown in Figure 3d, which are found to saturate at $|\xi_{DL}| \approx 1.1$ and $|\xi_{DL}| \approx 2.25$, respectively, as the $Bi_xTe_{1-x}$ layer gets thicker. These values are much greater than those obtained from W-based ($\xi_{DL}^W = -0.30 \pm 0.02$) and Pt-based ($\xi_{DL}^{Pt} = 0.08 \pm 0.01$) control samples. Moreover, this thickness depedence of $|\xi_{DL}|$ can be well-described by a bulk spin diffusion model with the spin diffusion length of $Bi_xTe_{1-x}$ $\lambda_s^{BiTe} \approx 4.1nm$. Our results therfore reveal that the non-epitaxial $Bi_xTe_{1-x}$ heterostructures can possess giant charge-to-spin conversion efficiencies originated from the bulk spin-orbit interaction of $Bi_xTe_{1-x}$, which is tentatively attributed to the SHE therein. This further indicates that the mechanism of spin generation in non-epitaxial chalcogenides can be drastically different from those observed in epitaxial TIs with TSS. Nevertheless, the post-annealing $Bi_xTe_{1-x}$ possesses a higher DL-SOT efficiency, which is possibly due to (1) the rougher interface between $Bi_xTe_{1-x}$ and Pt or (2) the improved crystallinity of $Bi_xTe_{1-x}$ may result in an additional TSS contribuion to the efficiency. Also note that the DL-SOT efficiency evaluated from hysteresis loop shift measurement is lower than that from harmonic Hall voltage measurement. It may due to the extra thermal effect included in harmonic Hall voltage measurement [26], whereas hysteresis loop shift measurement can avoid the additional contribution from that. Lastly, it is also important to note that several recent reports also support our observation that the SOT originated from non-epitaxial Bi-based materials has strong buffer layer thickness dependence and is mainly contributed to the bulk effect and intrinsic mechanism [33-34].

From above measurements, we verify that the non-epitaxial $Bi_xTe_{1-x}$ layers indeed has large DL-SOT efficiencies. To further demonstrate its efficacy, we perform current-induced switching measurements on as-deposited and post-annealing $Bi_xTe_{1-x}(15)/Pt(2)/Co(0.5)/Pt(2)$ Hall-bar devices. Pulsed currents are injected into current channel of the Hall-bar devices and magnetization switching can be monitored by the change of anomalous Hall voltage. Again, a small in-plane field $H_x$ is required to overcome the DMI effective field and to achieve deterministic SOT switching through chiral domain wall motion [35]. Representative switching results are shown in Figure 4a, b. We further summarize the critical switching currents vs. current pulse widths for both as-deposited and annealed samples in Figure 4c, d. For a spin-torque-driven magnetization switching scenario, the critical switching current $I_c$ is related to the applied current pulse width $t_{pulse}$ through [36]

$$I_c = I_{c0}\left[1 - \frac{1}{\Delta}\ln\left(\frac{t_{pulse}}{\tau_0}\right)\right], \tag{4}$$

where $I_{c0}$ is the zero-thermal critical switching current, $\Delta \equiv U/k_B T$ is the thermal stability factor ($U$ is the energy barrier between the magnetization-up and magnetization-down states), and $1/\tau_0 \approx 1$ GHz ($\tau_0 \approx 1$ ns) is the attempt rate for thermally-activated switching [37]. Based on the switching data shown in Figure 4c, d, we estimate $I_{c0} \approx 9.36$ mA with $\Delta \approx 33.3$ for the as-deposited Bi$_x$Te$_{1-x}$(15)/Pt(2)/Co(0.5)/Pt(2) device and $I_{c0} \approx 3.55$ mA with $\Delta \approx 32.3$ for the post-annealing device. After considering the device geometry, the zero-thermal critical switching current densities for as-deposited and post-annealing devices are found to be $|J_{c0}| \approx 1.73 \times 10^{10}$ A/m$^2$ and $|J_{c0}| \approx 6.54 \times 10^{9}$ A/m$^2$, respectively. The apparent DL-SOT efficiency can therefore be estimated from the direct current-induced switching results by considering $I_{c0}$ and the coercive field $H_c$ through [30]

$$\tilde{\xi}_{DL} = \frac{2e}{\hbar}\left(\frac{2}{\pi}\right)\mu_0 M_s t_{Co} w t_{Bi_x Te_{1-x}}\left(\frac{\rho_{Pt/Co/Pt} t_{Bi_x Te_{1-x}} + \rho_{Bi_x Te_{1-x}} t_{Pt/Co/Pt}}{\rho_{Pt/Co/Pt} t_{Bi_x Te_{1-x}}}\right)\left(\frac{H_c}{I_{c0}}\right), \tag{5}$$

where $H_c$ of the as-deposited (annealed) Bi$_x$Te$_{1-x}$ device is 27 Oe (12 Oe). By further taking the spin transmission factor into account, the intrinsic DL-SOT efficiencies are calculated to be $|\xi_{DL}| \approx 0.68$ for as-deposited and $|\xi_{DL}| \approx 0.82$ for post-annealing $Bi_xTe_{1-x}$. These values are fairly consistent with our characterization results, which support the conclusion that the sputtered, non-epitaxial $Bi_xTe_{1-x}$, either as-deposited or annealed, has a giant DL-SOT efficiency. However, it is interesting to note that although the switching current density of the $Bi_xTe_{1-x}$(15)-based device is indeed much lower than that of the W(2)-based device ($1.65 \times 10^{11}$ A/m$^2$, see Supporting Information S4), the switching currents are actually quite comparable in both systems. This indicates that for device applications, chalcogenides with even greater DL-SOT efficiency and shorter spin diffusion length are desirable.

Several remarks can be made based on our observations. First, the DL-SOT stemming from the non-epitaxial $Bi_xTe_{1-x}$ is comparable to those from epitaxially-grown TIs. This indicates that one does not need to struggle with preserving the TSS to achieve efficient SOT switching. It is very likely that using resistive materials such as amorphous and poly-crystalline Bi-based chalcogenides can already gain high SOT efficiencies through their bulk spin-orbit interactions, such as the SHE. Second, from both loop shift measurement and current-induced switching measurement, we verify that the DL-SOT from the SHE of $Bi_xTe_{1-x}$ can indeed control the magnetization of the Co layer. By annealing the $Bi_xTe_{1-x}$ layer, DL-SOT efficiency can also be improved. Third, the sign of the SHE-induced SOT acting on the Co layer is opposite to those reported for Bi-based TIs. This is perhaps due to (1) the difference in Bi concentration (which will affect the charge carrier type) while compared to other studies using epitaxial chalcogenides [38-39] or (2) the difference in terms of spin generation mechanism (bulk vs. TSS).

## 3. CONCLUSION

In summary, we systematically analyze the SOTs from the $Bi_xTe_{1-x}$-based magnetic heterostructures. Through harmonic voltage measurements and hysteresis loop shift measurements, we conclude that non-epitaxial $Bi_xTe_{1-x}$ has a large DL-SOT efficiency at room temperature, $|\xi_{DL}|>1$, which is much greater than those from HM-based heterostructures and comparable to those from other TI-family crystalline materials such as $Bi_xSe_{1-x}$, $Bi_xSb_{1-x}$, and $(BiSb)_2Te_3$. We further demonstrate current-induced switching in these $Bi_xTe_{1-x}$-based devices. We prove that a thermally-stable perpendicular magnetization (Co in Pt/Co/Pt) can be efficiently switched by means of the spin current generated from non-epitaxial $Bi_xTe_{1-x}$. Although the $Bi_xTe_{1-x}$ we prepared is not single-crystalline and therefore should not preserve TSS, it can still achieve high level DL-SOT efficiency, which is presumably originated from a bulk spin-orbit interaction contribution. Our work therefore paves the way of potential future research on non-epitaxial chalcogenide spintronic devices.

## MATERIALS AND METHODS

**Material Growth** All magnetic heterostructures are prepared by magnetron sputter depositions with base pressure $< 3\times 10^{-8}$ Torr. The $Bi_xTe_{1-x}$ thin films are deposited from a stoichiometric $Bi_2Te_3$ target and Pt/Co/Pt trilayers are deposited from Pt and Co targets. The magnetic heterostructures were directly deposited onto thermally oxidized silicon substrates. DC power is used to deposit all materials with 3 mTorr of Ar working pressure. No further thermal treatment is required to obtain PMA of the Pt/Co/Pt layer.

**TEM Characterization** High-Resolution TEM investigations are performed in a FEI Tecnai F20 at 200 kV. Cross-section foils are prepared using a focused-ion beam system (Helios NanoLab 600i). It should be noted that electron-beam deposition of platinum on the surface is necessary to protect the heterostructures.

**Device Fabrication** Hall-bar devices are patterned using standard photolithography and subsequent Ar ion-milling, with lateral dimensions of $5\,\mu m \times 60\,\mu m$. Ta(50 nm)/Pt(10 nm) are deposited as contact pads for electrical measurements.

**Magnetic Properties Characterization** Anomalous Hall voltage measurements are performed to confirm the PMA of our samples. The coercivity of the as-deposited and the post-annealing samples are around 27 Oe and 12 Oe, respectively. In-plane and out-of-plane hysteresis loops are measured by a vibrating sample magnetometer (VSM). The saturation magnetization of the Co layer in a Pt/Co/Pt structure is determined to be $M_s = 1414 \pm 120$ emu/cm$^3$ with no magnetic dead layer.

**Electrical Measurements** Harmonic voltage measurements are performed by measuring anomalous Hall voltages from the above-mentioned Hall-bar devices. A Signal Recovery 7265 lock-in amplifier is used to provide ac current and detect corresponding harmonic Hall voltages. A homebuilt electromagnet is used to apply magnetic fields along both $x$- and $y$-directions. For loop shift measurements, a Keithley 2400 source meter is used to provide dc current and a Keithley 2000 multimeter is used to measure the Hall voltage. The same homebuilt electromagnet supplies fields along $x$- and $z$-directions. For current-induced switching measurements, pulsed currents with pulse width between 0.01 s to 1 s are sent from the source meter to the device.

**SUPPORTING INFORMATION**

Supporting Information of this article are available online.

Section S1. EPMA analysis of the sputtered Bi$_x$Te$_{1-x}$ thin film

Section S2. Resistivity measurements

Section S3. Spin diffusion length of Pt spacer layer

Section S4. Control samples (W, Pt) SOT measurements

Figure S1. Resistivities of as-deposited Bi$_x$Te$_{1-x}$, Pt and Co

Figure S2. The spin diffusion length of the Pt spacer layer

Figure S3. Control samples based on W and Pt

Table S1. The EPMA analysis result of the as-deposited and post-annealing Bi$_x$Te$_{1-x}$ thin films

**Acknowledgements**

**Funding:** This work is supported by the Ministry of Science and Technology of Taiwan (MOST) under grant No. MOST-105-2112-M-002-007-MY3, No. MOST-108-2636-M-002-010, and No. MOST-107-3017-F-002-001 and by the Center of Atomic Initiative for New Materials (AI-Mat) and the Advanced Research Center of Green Materials Science and Technology, National Taiwan University from the Featured Areas Research Center Program within the framework of the Higher Education Sprout Project by the Ministry of Education (MOE) in Taiwan under grant No. NTU-108L9008 and No. NTU-108L9006. We thank Tsao-Chi Chuang and Hsia-Ling Liang for the support on VSM and AFM measurements.

**Author Contributions:** T.-Y. C. and C.-W. P. prepared the samples and performed hysteresis loop shift measurements. T.-Y. T. performed harmonic voltage measurements. W.-B. L. prepared and performed control measurements on conventional metallic heterostructures. C.-T. W and H.-W. Y. prepared the TEM samples and performed TEM investigations. C.-F. P. proposed and supervised the study.

**Competing interests:** The authors declare no competing financial interests.

**Data availability:** All data needed to evaluate the conclusions in the paper are present in the paper and/or the Supporting Information. Correspondence and requests for materials should be addressed to C.-F. P.

**REFERENCES AND NOTES**
(1) Wang, Y.; Deorani, P.; Banerjee, K.; Koirala, N.; Brahlek, M.; Oh, S.; Yang, H. Topological

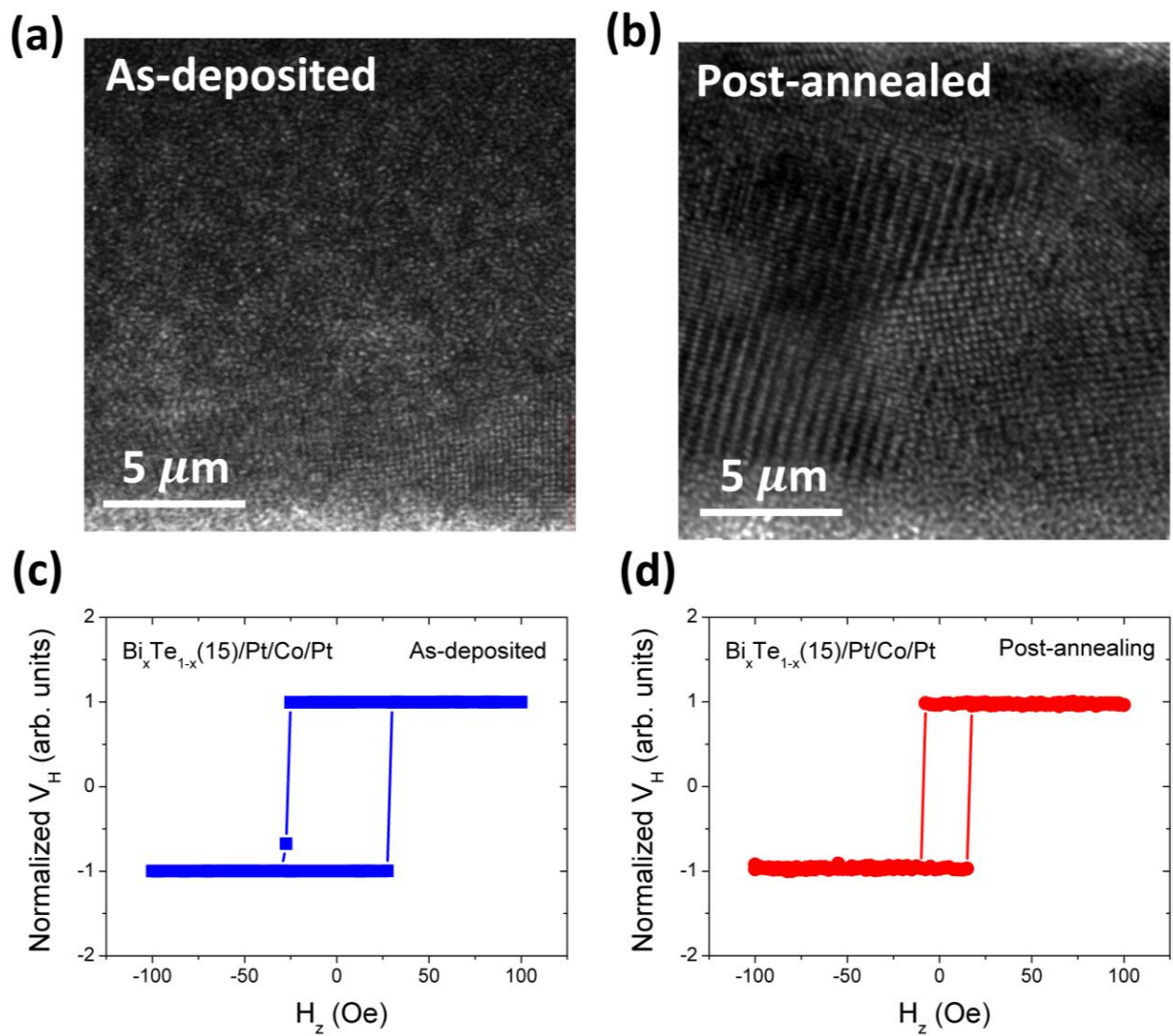

**Figure 1. Structural and magnetic properties of Bi$_x$Te$_{1-x}$-based heterostructures.** Cross-sectional HR-TEM imaging results from (a) an as-deposited and (b) a post-annealing 15 nm-thick Bi$_x$Te$_{1-x}$ thin film. The representative out-of-plane hysteresis loops as obtained by anomalous Hall voltage measurements of (c) an as-deposited and (d) a post-annealing Bi$_x$Te$_{1-x}$(15)/Pt(2)/Co(0.5)/Pt(2) heterostructures.

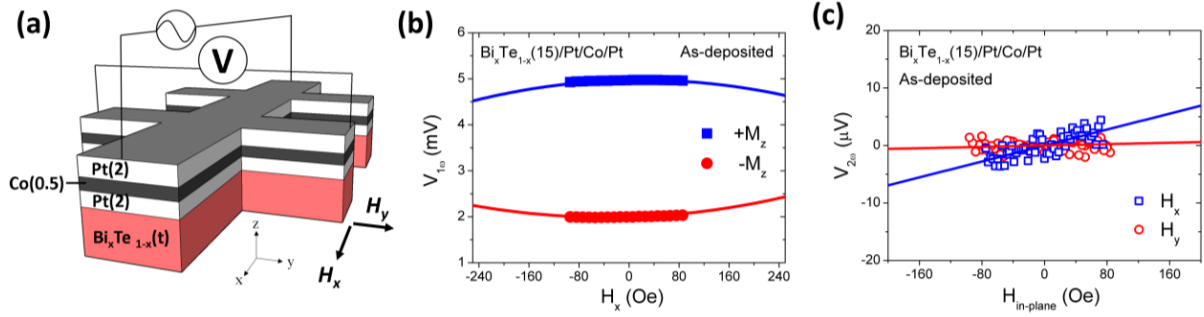

**Figure 2. Harmonic voltage measurements on an as-deposited Bi$_x$Te$_{1-x}$(15)/Pt(2)/Co(0.5)/Pt(2) Hall-bar device.** (a) Schematic illustration of a Hall-bar device for harmonic voltage measurement. $H_x$ ($H_y$) represents the applied longitudinal (transverse) in-plane field. Representative (b) first harmonic and (c) second harmonic Hall voltage results from an as-deposited Bi$_x$Te$_{1-x}$(15)/Pt(2)/Co(0.5)/Pt(2) Hall-bar device. The solid lines represent fits to extract curvatures (for $V_{1\omega}$) and slopes (for $V_{2\omega}$) to be used in equation. (1).

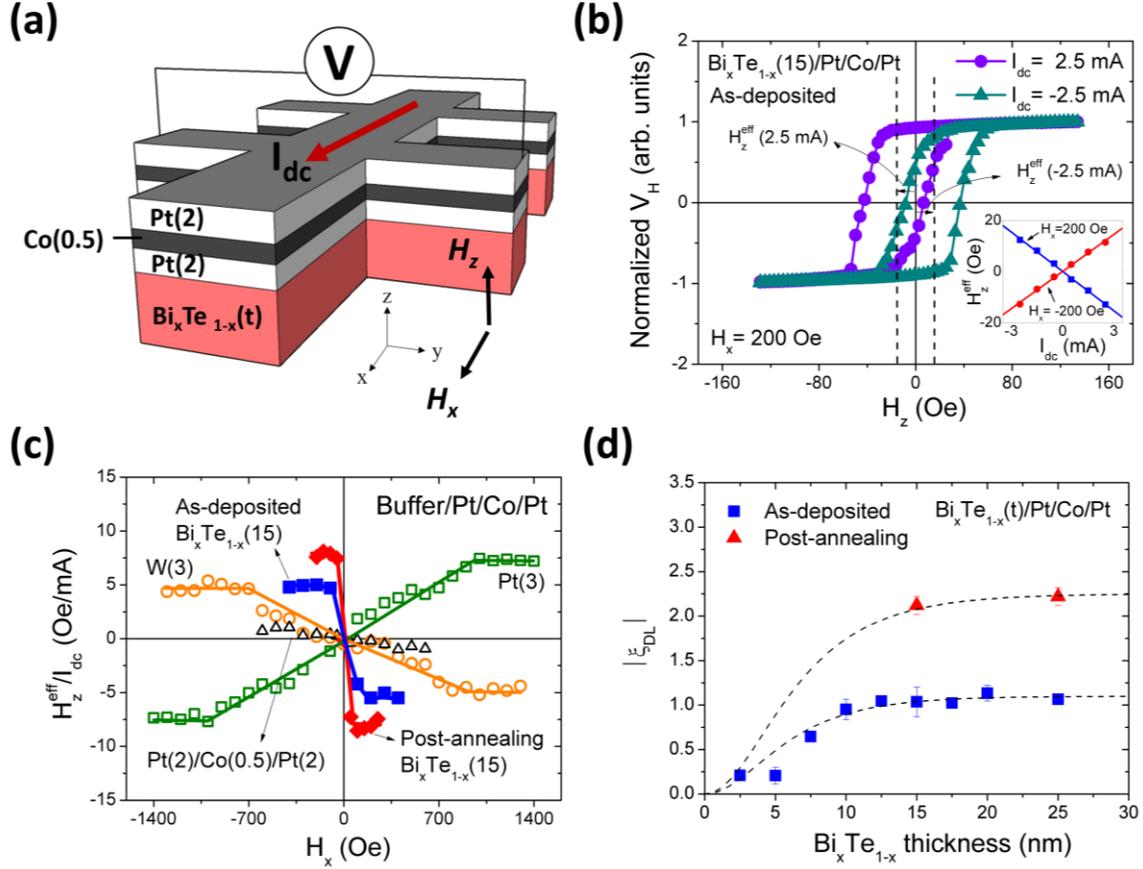

**Figure 3. Determination of DL-SOT efficiencies from Bi$_x$Te$_{1-x}$-based magnetic heterostructures.** (a) Illustration of a Hall-bar device for hysteresis loop shift measurement. $H_x$ and $H_z$ represent the applied in-plane and out-of-plane fields, respectively. (b) Representative out-of-plane hysteresis loops from an as-deposited Bi$_x$Te$_{1-x}$(15)/Pt(2)/Co(0.5)/Pt(2) Hall-bar device with different $I_{dc}$ and external in-plane field $H_x = 200\,\text{Oe}$. The inset shows out-of-plane current-induced effective field $H_z^{\text{eff}}$ as a function of $I_{dc}$ with $H_x = \pm 200\,\text{Oe}$. (c) $H_z^{\text{eff}}$ per $I_{dc}$ as functions of $H_x$ for heterostructures with different buffer layers. (d) Bi$_x$Te$_{1-x}$ thickness dependence of DL-SOT efficiencies for both as-deposited and post-annealing Bi$_x$Te$_{1-x}$-based heterostructures. The dashed lines represent fits to a spin diffusion model with spin diffusion lengths $\lambda_s^{\text{as-deposited}} \approx 4.1$ nm and $\lambda_s^{\text{post-annealing}} \approx 4.6$ nm.

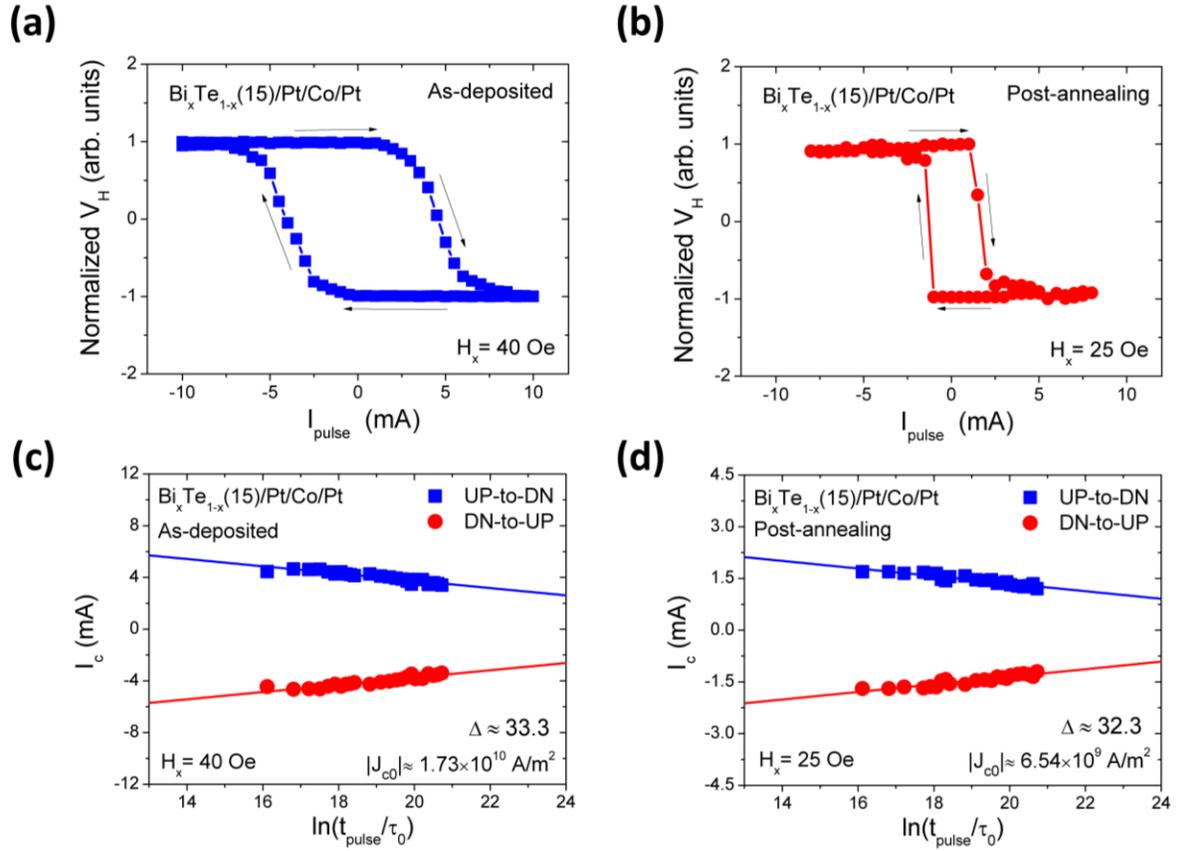

**Figure 4. Current-induced SOT switching results of an as-deposited and a post-annealing Bi$_x$Te$_{1-x}$(15)/Pt(2)/Co(0.5)/Pt(2) device.** Representative current-induced switching loops of (a) an as-deposited Bi$_x$Te$_{1-x}$(15)/Pt(2)/Co(0.5)/Pt(2) Hall-bar device and (b) an post-annealing Bi$_x$Te$_{1-x}$(15)/Pt(2)/Co(0.5)/Pt(2) Hall-bar device. Current pulse width $t_{pulse}$ dependence of critical switching current $I_c$ of (c) an as-deposited Bi$_x$Te$_{1-x}$(15)/Pt(2)/Co(0.5)/Pt(2) Hall-bar device and (d) a post-annealing Bi$_x$Te$_{1-x}$(15)/Pt(2)/Co(0.5)/Pt(2) Hall-bar device. Solid lines represent linear fits to the experimental data.

**Table of Contents (TOC) graphic**

For Table of Contents Only

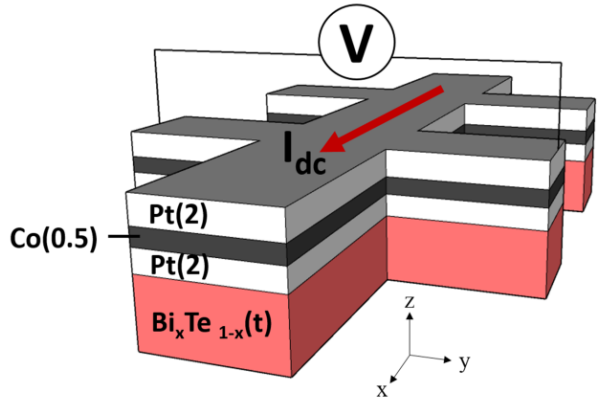 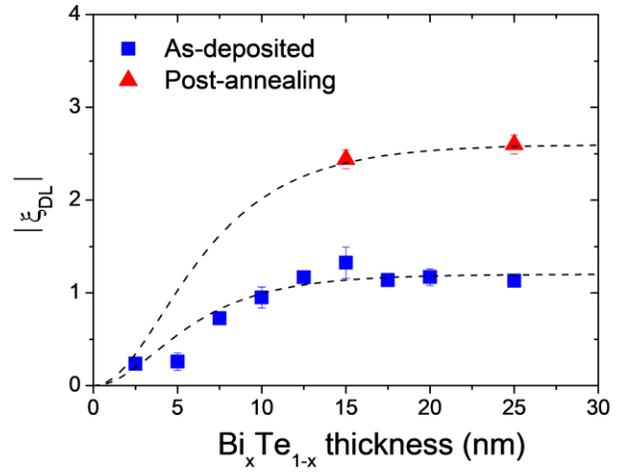